\documentclass{aa}
\usepackage{graphicx}
\usepackage{natbib}
\usepackage{amssymb}
\usepackage{amstext}
\usepackage{latexsym} 
\usepackage[sumlimits]{amsmath}
\newcommand{\Ha}{H$\alpha$}			
\newcommand{\NII}{[N{\sc ii}]}			
\newcommand{\HII}{H{\sc ii}}			
\newcommand{\HI}{H{\sc i}}			
\newcommand{\Msolar}{\mbox{\,$M_\odot$}}        

\begin{document}
   \title{The H$\alpha$ Galaxy Survey \thanks{
Based on observations made with the Jacobus Kapteyn Telescope operated 
on the island of La Palma by the Isaac Newton Group in the Spanish 
Observatorio del Roque de los Muchachos of the Instituto de Astrof\'\i sica 
de Canarias
   }}
   \subtitle{V. The star formation history of late-type galaxies}
   \author{Phil~A.~James,
    Matthew~Prescott and Ivan~K.~Baldry
    } 
          \offprints{P. A. James} 
          \institute{Astrophysics Research
	  Institute, Liverpool John Moores University, Twelve Quays
	  House, Egerton Wharf, Birkenhead CH41 1LD, UK \\
	  \email{paj@astro.livjm.ac.uk} 
	  }
          \date{Received ; accepted }

\abstract
{}
{This study of 117 low-redshift Im and Sm galaxies
investigates the star formation rates of late-type
galaxies, to determine whether they are quasi-continuous or 
dominated by bursts with quiescent 
interludes.} 
{We analyse the distribution of star formation 
timescales (stellar masses/star formation rates) for the entire
sample, and of gas depletion timescales for those galaxies with
gas mass measurements.}  
{We find that, on average, the late-type galaxies studied could
have produced their total stellar masses by an extrapolation of their 
current star formation activity over a period of just under a Hubble time.
This is not the case for a comparison sample of earlier-type galaxies, even 
those with disk-dominated
morphologies and similar total stellar masses to the late-type galaxies.
The earlier-type galaxies are on average forming their stars more slowly at 
present than the average rate over their past histories.
No totally quiescent Im or Sm galaxies are found, and although some 
evidence of intrinsic variation in the star formation rate with time is found,
this is typically less than a factor of 2 increase or decrease
relative to the mean level.
The Im and Sm galaxies have extensive gas reservoirs and can maintain star 
formation at the current rate for more than another Hubble time.  The average 
spatial distribution of star formation in the Im galaxies, and to a lesser 
extent the Sm galaxies, is very similar to that of the older stellar 
population traced by the red light. }
{Late type, bulge-free galaxies have a predominantly continuous mode
of star formation, and could have assembled their stellar masses through
continued star formation over a Hubble time with the currently-observed rate 
and spatial distribution.  There is little evidence in this sample of 
predominantly isolated field galaxies of significant star formation through
brief but intense starburst phases.}


\authorrunning{James et al.}
\titlerunning{H$\alpha$ Galaxy Survey. V.}
\maketitle
%
\section{Introduction}

Late-type galaxies of type Im and Sm are pure `blue sequence' star
forming systems, with little or no old bulge component, and are the
most numerous low-luminosity galaxy type in the field environment.
The star formation (SF) histories of these galaxies are of interest
for several reasons.  Low luminosity galaxies typically have shallow
potential wells, and are thus more susceptible to the effects of
internal feedback from stellar winds and supernovae \citep{deke86,deke03,
stin06}.
They are also likely to be affected by external factors such as the
intergalactic ultraviolet radiation field \citep{some02,greb04} which can
suppress SF in the lowest mass galaxies.  There is also significant
current interest in the fate of dwarf galaxies falling into galaxy
groups or clusters from the field environment \citep{maye01}; disentangling the
effects of environment on such galaxies requires a good knowledge of
the range of SF activity exhibited by isolated systems.  
In order to
address such statistical questions, large numbers of galaxies should
be investigated.  This is particularly true for the lowest luminosity
systems, for which SF indicators such as \Ha\ flux (used in the
present paper) may be substantially affected by statistical
fluctuations \citep{weil01}.

The present study looks at SF activity of 42 Sm and 75 Im galaxies
from the H$\alpha$ Galaxy Survey \citep[H$\alpha$GS;][]{paper1}. The
analysis centres on the SF timescale for each of the galaxies studied,
where this timescale is defined as the ratio of the total stellar mass
to the current SF rate. This method is thus closely related to the 
stellar birthrate parameter pioneered by \cite{kenn94}. The distribution of 
SF timescales provides
a test of whether SF is continuous, as found observationally by
e.g. \citet{gall84} and \citet{vanz01}, or in temporally separated
bursts.  The latter have been predicted by theoretical studies such as
the Stochastic Self-Propagating Star Formation models of
\citet{gero80}, and by recent simulations carried out by
\citet{stin07}. Observational evidence of burst-dominated SF in
dwarfs has been found by \citet{almo98} in a study of late-type Virgo
cluster dwarfs, and by \citet{bara06} who looked at dwarf galaxies
over to redshift range 0.02 - 0.25.
In addition, studies based on abundance ratios and comparison with
chemical evolution models have found evidence of burst-dominated SF histories
in Local Group dwarfs, and in blue compact galaxies \citep{lanf03, recc06}.

The present study is the largest sample of local late-type galaxies
studied using \Ha\ imaging.  The 117 galaxies of the present sample
constitute $\sim$55\% of the known Sm and Im galaxies satisfying the
selection criteria of the \Ha GS sample, and they are all of the
  observed galaxies of these types in the \Ha GS study. Thus we can
with confidence extrapolate any results found to the general
population of late-type field galaxies in the local Universe; there is
no reason to suspect any systematic differences between the 55\% of
galaxies observed and the remaining 45\%.

We also carry out further tests of SF activity in \Ha GS galaxies,
using subsets of objects for which other data are available.  The
2MASS database \citep{jarr03} provides homogeneous near-infrared (NIR)
photometry for 201 out of the full sample of 334 \Ha GS
galaxies, over all morphological types.  We calculate our own $J$ and
$K$ total magnitudes for these galaxies, based on the 2MASS aperture
photometry, and hence derive stellar mass estimates as a check on
those derived from our own $R$-band photometry.  The $R - K$ and $J -
K$ total colours can also be compared with the predictions of stellar
population synthesis models, giving a consistency check on our
conclusions about the SF activity in the late-type galaxies and the
mass-to-light (M/L) ratios assumed in the calculation of the SF timescales.

Gas masses derived from \HI\ observationss are available for 94 
galaxies in the \Ha GS sample from the
Westerbork observations of neutral Hydrogen in Irregular and SPiral
galaxies (WHISP) survey \citep{swat02}.  This enables a calculation of
the gas depletion times for galaxies of different types, which gives a
further consistency check on our conclusions.

Finally, we compare the spatial distributions of forming and old
stellar populations in the late-type galaxies, from the \Ha\ and
$R$-band imaging respectively, to test whether the overall
structures of galaxies are consistent with their having been built up
by SF activity distributed like that seen currently.

The structure of the current paper is as follows.  Section
\ref{sec:data} contains a description of the galaxy sample, the data
used and the calculation of stellar masses for all galaxies.
Section \ref{sec:sft} outlines the calculation of the timescale 
required for the formation of the total stellar mass of each
galaxy, given the current SF rate derived from \Ha\ emission.
Section \ref{sec:gdt} similarly looks at the timescale for the 
conversion of the current gas reservoirs into stars, assuming 
continued SF at the current rate. 
These conclusions are combined in 
section \ref{sec:bursts} in an analysis of the `burstiness' of 
SF, i.e. the extent to which the SF rate of individual galaxies
varies about the mean level.
Section \ref{sec:othertests} contains an analysis of the extent to which 
the conclusions depend on specific stellar population synthesis models 
used, and whether the initial assumptions used to calculate the 
stellar M/L ratio are consistent with the final conclusions. 
Section \ref{sec:sum} is a summary of the main conclusions.

\section{Photometric data and derived quantities}
\label{sec:data}

The \Ha\ Galaxy Survey (\Ha GS) is a study of the SF properties of
galaxies in the local Universe, using fluxes in the \Ha\ line to
determine the total rates and distributions of SF within the selected
galaxies.  The observations cover 334 galaxies, all of which
were observed with the 1.0 metre Jacobus Kapteyn Telescope (JKT), part
of the Isaac Newton Group of Telescopes (ING) situated on La Palma in
the Canary Islands.  The selection and the observation of the sample
are discussed in \citet{paper1}, hereafter Paper I, but to summarise, 
galaxies were selected from the Uppsala General Catalogue of Galaxies
\citep{nils73} with diameters between 1\farcm7 and 6\farcm0, recession
velocities less than 3000~km~s$^{-1}$, and Hubble types from S0a -- Im
inclusive.  
The Virgo cluster region was excluded from selection, so this is 
predominantly a field galaxy sample. Approximately 50\% of the
galaxies (and almost all of the Sm and Im types) are within a 15~Mpc 
radius from the Milky Way but, 
given the diameter upper limit,
no Local Group objects are included.  Thus the late-type (Sm and Im) sample 
closely resembles that of \citet{vanz01} in terms of environment, distance
and luminosity, but differs from samples studied
via analysis of resolved stellar populations, where the galaxies tend to be
nearby low-surface-brightness extreme dwarfs.
All \Ha GS $R$-band magnitudes and \Ha\ fluxes used here
are corrected for Galactic extinction according to the methods of
\citet{schl98}.

\subsection{Stellar masses}

Stellar masses were initially derived from the $R$-band total
magnitudes of each of the \Ha GS galaxies, with the 
M/L ratio being derived from the models of \citet{bell01} and a
type-dependent $(B-R)$ colour taken from \citet{bell00}.
Table~\ref{tbl:col_ml} lists the mean adopted colours and the
resulting M/L ratios as a function of galaxy type.  \citet{bell01}
base their models on a Salpeter IMF but scale it
by a factor 0.7 to avoid violating dynamical disk mass limits
(`diet Salpeter').  This
also brings the M/L ratios closer to the \citet{chabrier03} and
\citet{kroupa01} IMFs which contain fewer low-mass stars than implied
by a single power law.

Errors on the $R$-band-derived stellar masses result from photometric
errors (typically 10\% for our $R$-band photometry), internal
extinction errors (taken as 50\% of the \Ha\ extinction errors,
explained in section 2.2 below) and uncertainties in the M/L ratios
from modelling, which we took as 40\%.  These stellar masses thus have
errors in the range 43 - 48\%, with the larger values corresponding to
the brighter galaxies.

\begin{table*}
\begin{center} 
\begin{tabular}{c|ccc}
\hline
\hline
 Galaxy T-type & $(B-R)$ & (M/L)$_R$ &  (M/L)$_K$    \cr
\hline
 0 - 2  & 1.50 & 2.86 & 0.80 \cr
 3 - 4  & 1.30 & 1.93 & 0.65 \cr
 5 - 6  & 1.25 & 1.75 & 0.62 \cr
 7 - 9  & 1.00 & 1.07 & 0.47 \cr
  10    & 0.74 & 0.65 & 0.36 \cr
\hline
\end{tabular}
\caption[]{Mean colours and M/L ratios in the $R$ and $K$
bands as a function of galaxy T-type.}
\label{tbl:col_ml}
\end{center}
\end{table*}

A check on the stellar masses of many galaxies was provided by the
availability of 2MASS NIR photometry for 201
galaxies in the \Ha GS sample.  The NIR has significant
advantages over the optical in this context; extinction corrections
are much smaller, and in general M/L variations are a weaker
function of SF history, in the redder passbands.  However, there are
limitations of the 2MASS database for the current study, as many of
the latest-type galaxies have low surface brightnesses in the NIR and
are not detected by 2MASS.  Care was needed in allocating total $J$
and $K$ magnitudes to those galaxies that were in the 2MASS
database. In many cases, the published 2MASS extrapolated magnitudes,
$J_{\rm ext}$ and $K_{\rm ext}$, were found to be too faint by several
tenths of a magnitude, e.g. by comparison with 2MASS circular aperture
photometry.  This is due to a known problem with the 2MASS
extrapolation algorithm, which can lose flux for extended objects of
low surface brightness (T. Jarrett, private communication).  For all
affected galaxies, we derived our own $J_{\rm tot}$ and $K_{\rm tot}$
magnitudes, by matching 2MASS circular aperture photometry to our own
$R$-band growth curve, which was derived from matched circular
apertures, and assuming an identical $\Delta m$ between the largest
aperture and the asymptotic total magnitude in the $R$, $J$ and $K$
passbands.  This assumes the $R-J$ and $R-K$ colour gradients are
small in the outer regions of these galaxies.

After applying these methods, we have homogeneous $J_{\rm tot}$ and
$K_{\rm tot}$ magnitudes for 201 of the \Ha GS galaxies,
more than 50\% of the total, although this is biased towards the
brightest and highest surface brightness types.  In particular, only 8
Sm and 13 Im galaxies have total 2MASS magnitudes, and these are very
much weighted towards the brightest examples of these types.

\begin{figure}
\centering
\rotatebox{0}{
\includegraphics[height=7.5cm]{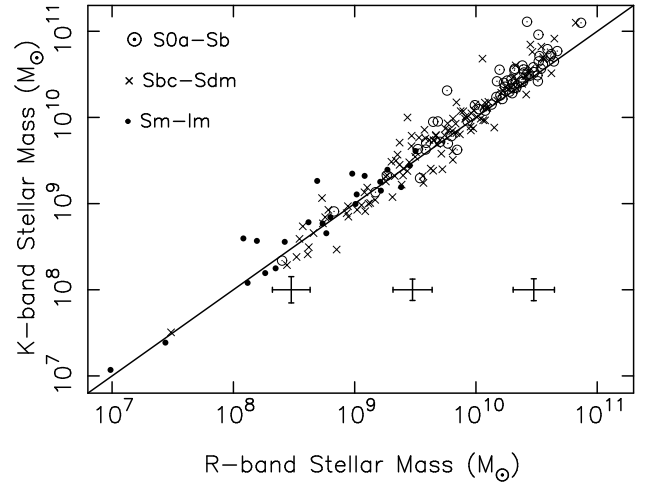}
}
\caption{The $K$-band derived stellar mass 
versus the $R$-band derived stellar mass 
for the 201 galaxies with 2MASS photometry.
Three representative error bars are shown, corresponding to galaxies
with masses $<$10$^9M_{\odot}$;
10$^9M_{\odot}$ - 10$^{10}M_{\odot}$; and
$>$10$^{10}M_{\odot}$, respectively.}

\label{fig:masscomp}
\end{figure}

Figure~\ref{fig:masscomp} provides a check on the reliability of the
$R$-band stellar mass estimates, using the 2MASS $K_{\rm tot}$
magnitudes described above.  The quantities plotted on the vertical
and horizontal axes are the individual galaxy stellar masses estimated
from the $K$- and $R$-band magnitudes respectively, using M/L ratios
from \citet{bell01} as in Table~\ref{tbl:col_ml}.  Stellar masses
derived from $K$-band photometry are subject to the same sources of
error as those from $R$-band photometry.  Although the formal errors
on the published 2MASS data are small, the total magnitudes used here
have significant uncertainties due to the extrapolation procedure
described above. This gives the largest errors (0.3~mag) for the faint
galaxies, falling to 0.2~mag for the brightest ones.  Extinction and
M/L modelling uncertainties are smaller than for the $R$-band,
$\sim$25\%.  Thus the total errors in $K$-band-derived stellar masses
are 42\% for the lower-luminosity galaxies, falling to 33\% for the
bright spirals.

Figure~\ref{fig:masscomp} shows a significant scatter
in the mass estimates from the two passbands, with individual galaxies
showing discrepancies as large as a factor of 3, in either sense.
The typical error bars shown in Fig.~\ref{fig:masscomp} are calculated 
from the error sources described above, and do not include contributions due to
galaxy distance uncertainties, since these affect both mass estimates equally
and do not contribute to the scatter.  

It can be noted from Fig.~\ref{fig:masscomp} that there is a small but
systematic offset in the sense that the $K$-band-derived masses are
slightly larger than those from the $R$-band luminosity.  This offset
is a factor of 1.12$\pm$0.08 in the mean for galaxies with masses
below 10$^9M_{\odot}$; 1.09$\pm$0.04 for galaxies between 10$^9$ and
10$^{10}M_{\odot}$; and 1.28$\pm$0.04 for galaxies more massive than
10$^{10}M_{\odot}$.  This may indicate a systematic offset inherent in
the M/L modelling, or under-correction for internal extinction which
would preferentially affect the brighter galaxies.  In any case, the
effect is small compared with other sources of error for the faint Sm
and Im galaxies with which the present paper is primarily concerned.
The $R$-band stellar masses were used in the rest of our analysis,
since these are available for the full sample of galaxies.

\subsection{Star formation rates}


The SF rates quoted in \citet{paper1} were determined from the
\Ha\ fluxes using the \citet{kenn98} conversion factor for a
\citet{salp55} stellar initial mass function (IMF) from 0.1 to
100\Msolar.  However, as explained above, several studies have now
concluded that this IMF results in an unrealistically high total
stellar mass, and one solution is to multiply all stellar masses and
SF rates by 0.7.  Since this is done for the stellar
mass estimates used here, for consistency we scale the SF rates from
\citet{paper1} by the same factor of 0.7.  We assume that the IMF is
universally applicable. If the IMF varies between galaxies, as
suggested by \citet{hove07}, there could be substantial variations in
the conversion of \Ha\ flux to SF rate and colour to stellar mass that
will affect the timescale determinations.

There are several contributors to the errors on derived SF rates.  The
\Ha\ fluxes have errors which are dominated by uncertainties in
continuum subtraction \citep{paper1}, with further contributions due
to filter throughput, sky background and photometric calibration
uncertainties.  In total, the \Ha\ flux errors for the lower
luminosity galaxies in which we are most interested here are of order
25\%, falling to 10\% for the brighter galaxies with strong
\Ha\ emission.  The other major sources of error in the calculated SF
rates arise from the adopted extinction corrections, and to a a lesser
extent the corrections for the \NII\ lines that lie in the bandpass of
the narrow band filters used.  Several sets of such corrections were
investigated, including those of \citet{kenn83}, a type-dependent
correction from \citet{paper2}, and corrections dependent on $R$-band
total magnitudes, following the prescription of \citet{helm04}.  The
latter were adopted for the present paper, and following an
intercomparison of the size of the extinction corrections from the
different methods, errors were assigned of 0.4~mag. for the brightest
spiral galaxies, falling to 0.26~mag for the fainter spirals and
0.16~mag for the faintest galaxies, including almost all of the Sm and
Im types.  This error corresponds to about 50\% of the size of the
applied extinction correction.  Combining the \Ha\ flux and extinction
errors, the overall errors on SF rates are 30\% for the faintest
galaxies, rising to 50\% for the brightest spirals, with the rise
being due to the larger, and hence more uncertain, extinction
corrections for larger galaxies.

\section{Analysis of star formation timescales}
\label{sec:sft}

The initial analysis involved the calculation of the time required to
form the total stellar mass of each galaxy studied, under the
assumption of continuous SF at the current rate.  This was given by
\begin{equation}
{\rm SF~timescale} = \frac{\rm stellar~mass}{\rm SFR} 
  \frac{1}{(1-{\cal R})}
\label{eqn:sf-timescale}
\end{equation}
where ${\cal R}$ is the mass recycled to the interstellar medium per
mass of stars formed. The factor ${\cal R}$ is taken to be 0.4
as per \citet{vanz01}. 
The metallicity dependence of ${\cal R}$ was investigated
using the BaSTI population synthesis model \citep{piet04} and 
it was found that the recycled fraction is
slightly larger at low metallicity than at solar metallicity.
However, the change in recycled fraction only leads to a 3.5\% change
in $(1-{\cal R})^{-1}$ for a 10 Gyr old population,
for a change in [Fe/H] of +0.06 to --1.49.  The change is smaller than
this at younger ages (e.g. 2\% at 1~Gyr).

Note that the recycling correction in
Eq.~\ref{eqn:sf-timescale} is appropriate because the calibrations of
M/L by \citet{bell01} are in relation to remaining stellar mass, which
is the more generally used definition of stellar mass, as opposed to
the integral of the SF rate.

The SF timescale enables the identification of quiescent galaxies, for
which this timescale will be much greater than a Hubble time, and
galaxies currently in a starburst phase, for which it will be much
less than a Hubble time.

\begin{figure}
\centering
\rotatebox{0}{
\includegraphics[height=7.5cm]{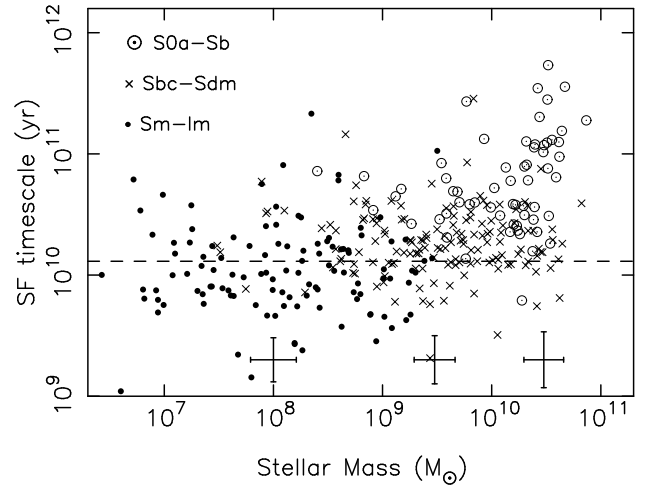}
}
\caption{The SF timescale as a function of galaxy stellar mass. 
The dashed line indicates the age of the Universe, 13.7~Gyr.
Representative error bars are shown for the same three mass ranges
as Fig. \ref{fig:masscomp}.  In this case, the horizontal error bar includes
an error due to galaxy distance uncertainties; the quantity plotted on the
vertical axis is distance-independent. 
}
\label{fig:sftpts}
\end{figure}

\begin{figure}
\centering
\rotatebox{0}{
\includegraphics[height=7.5cm]{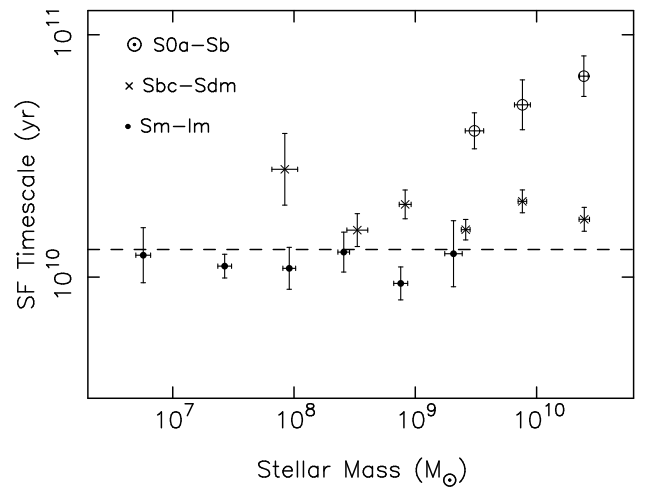}
}
\caption{Mean SF timescale as a function of galaxy stellar mass.
The dashed line indicates the age of the Universe, 13.7~Gyr.
Note the reduced vertical scale cf.\ Fig.~\ref{fig:sftpts}.}
\label{fig:sftbin}
\end{figure}

Figure~\ref{fig:sftpts} shows the SF timescale values for individual
galaxies in the \Ha GS sample, with stellar masses and internal
extinction corrections derived from $M_{R{,\rm tot}}$ magnitudes, as
explained in the previous section.  The different point styles
indicate galaxies in 3 different morphological type ranges: S0a - Sb,
Sbc - Sdm, and Sm - Im.  
Representative error bars are shown for the same three mass ranges
as Fig. \ref{fig:masscomp}.  In this case, the horizontal error bar 
additionally includes
an error due to galaxy distance uncertainties; the quantity plotted on the
vertical axis is distance-independent.  The distance errors, derived
from a comparison of our values from \citet{paper1} with estimates
in the literature, were taken as
20\% in distance (44\% in estimated luminosities and quantities derived 
from them) for galaxies nearer than 10~Mpc, which includes most of the Sm and Im
types.  For the more distant galaxies, the distance errors decrease to 10\%.

Some trends and differences as a function of
morphological type are already evident in Fig. \ref{fig:sftpts}, but are shown
more clearly in Fig.~\ref{fig:sftbin}, in which the individual galaxy
SF timescales from Fig.~\ref{fig:sftpts} are binned over ranges of
galaxy stellar mass.  In Fig.~\ref{fig:sftbin}, the quantity plotted
for each type/mass bin is
\begin{equation}
\frac{\Sigma_i ({\rm mass}_i)}{\Sigma_i ({\rm SFR}_i)} \frac{1}{(1-{\cal R})}
\label{eqn:timescale-averaging}
\end{equation}
rather than the mean of the individual timescale values, to minimise
the effect of outliers.  Figure~\ref{fig:sftbin} shows clearly that
the bulge-dominated early-type galaxies have high timescales, implying
that their current SF rates are significantly lower than would be
required to form the stellar mass within a Hubble time.  The late-type
Sm and Im galaxies, on the other hand, have uniformly short SF
timescales, with binned means in the range 9--13~Gyr. Thus these 
bulge-free galaxies could have formed their
stellar mass through continuous SF activity at or slightly below the
current rate. It is important to note that the mass ranges of the
different types overlap significantly, and the shorter SF timescales
can be seen to depend only on the lack of a bulge, and not on galaxy
mass.  Another interesting feature of Fig.~\ref{fig:sftpts} is that
the scatter in the late-type galaxy SF timescales is constant as a
function of galaxy mass, at least over the range studied, and shows no
sign of an increase at low galaxy masses and low SF rates, as would be
expected if statistical fluctuations were dominating the scatter.
This scatter will be analysed further in Section \ref{sec:bursts}.

\begin{figure}
\centering
\rotatebox{0}{
\includegraphics[width=0.5\textwidth]{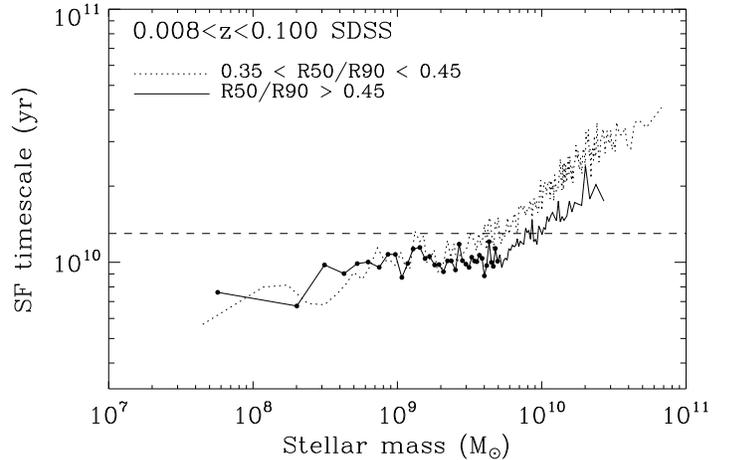}
}
\caption{Mean SF timescales derived from the SF rates and stellar masses of
  \citet{brinchmann04}.
The dashed line indicates the age of the Universe, 13.7~Gyr.
}
\label{fig:sdss}
\end{figure}

Star formation timescales of $\sim$10~Gyr were also found for the
field Magellanic irregulars studied by \citet{vanz01}, and for the
dwarf irregulars in the M~81 group study of \citet{kara07}; however,
\citet{skil03} found much longer timescales for dwarf irregular
galaxies in the Sculptor group, which they identified as a possible
transition dI - dE population. In all three studies, the methods
employed were very similar to the present work.  We also analysed SF
rates and stellar masses from the Sloan Digital Sky Survey (SDSS) as
derived by \citet{brinchmann04}. Taking the Data Release Four sample,
we selected galaxies with $0.008<z<0.1$, $17.5<r<17.75$ (near the
limit of the main galaxy sample to minimise aperture effects) and an
estimated SF rate. The data were then divided into low concentration,
predominantly late types ($R50/R90> 0.45$ in the $r$-band) and
intermediate concentration ($0.35 < R50/R90 <
0.45$). Figure~\ref{fig:sdss} shows a plot of the timescales in bins
of 100 galaxies with the average as per
Eq.~\ref{eqn:timescale-averaging} except a weight equal to $1/D^3$ was
applied to each galaxy where $D$ is the distance. This is because
within a narrow magnitude range, the volume over which a galaxy could
be observed is approximately proportional to $D^3$.  The timescales
are $\sim$10~Gyr except at high masses where the contribution of
earlier types comes in and at low masses where there are various
complications with SDSS data such as incompleteness.

To conclude this section, we can propose a simple model of galaxy SF
histories, consistent with above findings. We find evidence of two
modes of SF, one at a constant rate, and the other
dominated by an earlier burst.  The Sm and Im late-type galaxies
appear to be dominated by the first mode, whilst earlier type galaxies
have a combination of the two modes, with the early burst becoming
more dominant for earlier Hubble types and higher masses, at least
within the earliest types.

\section{Gas depletion timescales}
\label{sec:gdt}

\begin{figure}
\centering
\rotatebox{0}{
\includegraphics[height=7.5cm]{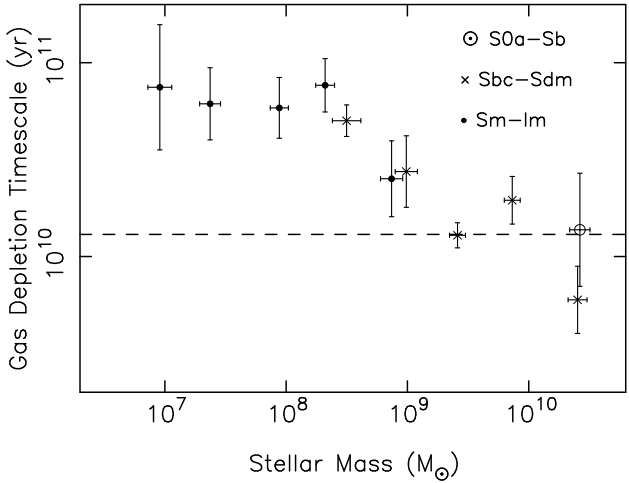}
}
\caption{Mean gas-depletion timescale plotted against galaxy stellar 
mass.  The dashed line indicates the age of the Universe, 13.7~Gyr.
}
\label{fig:gdtbin}
\end{figure}

Continued SF with no indication of a fall-off at the current epoch
implies a substantial reservoir of gas in the galaxies concerned. This
is confirmed in Figure~\ref{fig:gdtbin}, which shows the gas depletion
timescales for the 94 members of the current galaxy sample with
\HI\ masses provided by the Westerbork observations of neutral
Hydrogen in Irregular and SPiral galaxies (WHISP) survey
\citep{swat02}.  The \HI\ masses are multiplied by 2.3 to account for
helium and molecular gas; this factor was taken from \citet{meur06}
who base their calculation on H$_2$:\HI\ ratios from the study of
\citet{youn96}.  The resulting gas mass divided by the SF rate,
corrected as above for the recycling factor ${\cal R}$, gives the
depletion time.  In general, the \HI\ is more widely distributed
than the stellar component in galaxies; for example, \citet{swat02}
find the \HI\ extents to be typically 1.8 those of the stars, with 
significant scatter.  No correction was made for this effect, as even 
the outlying gas can be considered as ultimately available to fuel
star formation.

Errors in gas masses were attributed to two causes: errors in the
\HI\ measurements from the WHISP study, and uncertainty in the
conversion of \HI\ to total gas masses.  The former were taken from
\citet{swat02}, who quote a 15\% RMS difference between their flux
densities and those quoted in the literature for the same galaxies.
They also comment that there should be no problems with large-scale
flux being resolved out for galaxies as small as those studied here.  Thus
15\% was adopted as the error on \HI\ flux density, and hence on 
atomic gas mass.  The
conversion constant of 2.3 is significantly more uncertain; for
example, \citet{garn02} finds molecular gas fractions in many galaxies
to be generally smaller than those found by \citet{youn96}. We
adopt a 30\% error from this cause, giving 34\% overall errors in gas
masses.  These need to be added in quadrature to the SF rate errors,
to give errors on the gas depletion timescales of 45 - 61\%, the
largest values being for the brightest galaxies.

Figure~\ref{fig:gdtbin} shows that spiral galaxies can have short gas
depletion times, particularly at high masses, in spite of their low
specific SF rates implied by Figs. \ref{fig:sftpts} and
\ref{fig:sftbin}.  The Sm/Im galaxies on the other hand have ample gas
reservoirs to continue their current SF activity for much more than a
further Hubble time.  For those late-type galaxies with stellar masses
below 10$^9$~M$_{\odot}$, there is a marked plateau in gas depletion
times at 7$\times$10$^{10}$~years; these galaxies have
sufficient fuel to continue their SF for several Hubble times, even
without gas replenishment.  Thus there is no contradiction or
fine-tuning implied in the suggestion that the SF activity has
continued unabated to the present epoch.

\section{A determination of the `burstiness' of Sm/Im SF}
\label{sec:bursts}

We have argued in the previous sections that the properties of the
late-type galaxies are consistent, on average, with their having
assembled their stellar populations through SF at a constant rate that
is similar to that currently observed.  We now look in more detail
to see to what extent this can be the case for individual galaxies.

A first point to note is that no completely passive Sm or Im galaxy
is found in our sample of 117 objects.  In every case, detectable \Ha\
emission is found.  This is significant, as the present sample is a
substantial fraction of the total number of late-type dwarfs in the
local Universe; we have observed 55\% of the Sm and Im galaxies in the
UGC that satisfy the \Ha GS selection criteria.  Thus, unless we have
been extremely unlucky in our selection, passive Sm and Im field
galaxies must be extremely rare or non-existent.  Such passive
galaxies cannot be masquerading as dwarf ellipticals, as such objects
are rare in the field environment.  Some could evade detection by
fading below selection limits, but this would require the passive
phase to have continued for several Gyr to be a significant effect.
Thus we conclude that all galaxies that have the gas reservoirs to
support SF are in fact observed to form stars.  This same conclusion
was also reached by \citet{meur06}, who imaged 93 \HI-selected
galaxies in \Ha\, and detected line emission in all of them.  
In addition, the analysis by \citet{haines07}, using SDSS data, found
all of the $\sim600$ low-luminosity galaxies studied ($-18 < M_r <
-16$) in the lowest-density environments to have \Ha\ emission.

There is significant
scatter in the SF timescales plotted in Fig.~\ref{fig:sftpts}, which
is consistent with SF rates in individual galaxies varying by factors
of a few about the mean
rate.  Formally the scatter in SF timescale in Fig.~\ref{fig:sftpts} is
0.36 dex for types Sm and Im (cf. 0.26 dex for types Sbc to Sdm and 0.34 dex
for types S0a to Sb).  From the error analysis presented earlier, the 
expected errors on SF timescale are 52\% for the late-type galaxies, or 
0.18 dex; thus, it would apear that the bulk of the vertical scatter in   
Fig.~\ref{fig:sftpts} is due to intrinsic variations in the SF timescale,
and not just measurement error.

\begin{figure}
\centering
\rotatebox{0}{
\includegraphics[height=7.5cm]{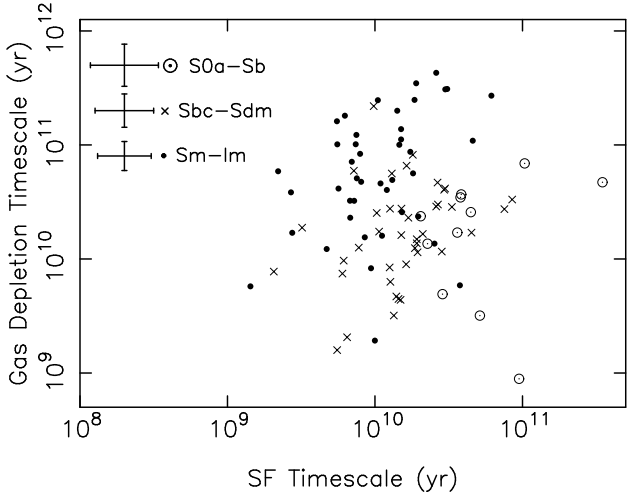}
}
\caption{Gas depletion timescale plotted against SF
timescale, for the 94 galaxies with \HI\ gas mass determinations.
Representative error bars are shown for galaxies in each of the
3 T-type bins.}
\label{fig:sftgdt}
\end{figure}

A further analysis of the burstiness of SF can be made using
a technique similar to the one applied by \citet{kara07} to the study of
galaxies in the M~81 group.  In their Fig.~4, they plot a parameter
F$_{\star}$, which is very similar to our SF timescale, against
P$_{\star}$, which is closely related to our gas depletion timescale
(with the main differences being that they normalise both parameters
explicitly by the Hubble time, and F$_{\star}$ is based on $B$-band
luminosity).  They find a strong correlation between these parameters,
which they interpret in terms of some galaxies being in starburst and
quiescent phases, which respectively shorten and lengthen the timescales
corresponding to these parameters.  Figure~\ref{fig:sftgdt} shows the
equivalent plot for the 94 \Ha GS galaxies with \HI\ gas mass
determinations.  Overall we find no correlation (correlation
coefficient --0.02) between the SF and gas depletion timescales when
considered across all galaxy types; however, for the 43 Sm and Im
galaxies there is a marginally significant correlation at the 
2.3 sigma level (correlation coefficient 0.33;
significance 0.028) in the expected sense that galaxies with longer SF
timescales also tend to have longer gas depletion timescales.  

We can use Fig.~\ref{fig:sftgdt} to quantify the burstiness of SF by
noting that an instantaneous increase or decrease in SF rate will move
points to the upper right or lower left, thus increasing the scatter
along this 45$^{\circ}$ axis relative to that along the perpendicular
axis.  The additional scatter along this axis, that is required 
to explain the 2.3 $\sigma$ correlation implies a variation in SF rate
by 0.247 dex, or a factor of 1.8 about the mean rate, for the Sm and
Im galaxies.  This modulation for our field sample is consistent with
the scatter in both the SF and gas depletion timescales, which
cannot be explained though measurement errors, and is substantially
lower than that required to explain the findings of \citet{kara07} for
the M~81 group. The stronger variation in SF rates that they find may
be due partly to the M~81 group environment being ideally suited to
strong inter-galaxy tidal interactions, and partly due to their sample
including more very low luminosity galaxies where the intrinsic
variations in SF rate may be more marked. \citet{noes07}, in a study
of the long-term SF trends in field galaxies, find a gradually declining 
SF rate, with strong SF bursts being uncommon in their sample of galaxies,
in good overall agreement with our findings.

\section{Other tests of continuous SF histories in field late-type dwarfs}
\label{sec:othertests}

In this section we perform two final tests of the consistency of
constant SF rates in Sm/Im galaxies with the observed galaxy
properties.  The first test addresses potential concerns over the use
of the \citet{bell01} calibration of the mean stellar M/L ratios based
on mean colours taken from the literature. This method raises several
questions: do the mean properties apply to the specific galaxies
observed here, how sensitive are the mass predictions to the
particular stellar synthesis models used by \citeauthor{bell01}, and
does our proposed SF history result in a population with a M/L ratio
consistent with that assumed in our calculation? In order to give
complete independence from the \citeauthor{bell01} models, we use use
the BaSTI population synthesis code \citep{piet04} for this test. This
code was used to generate stellar populations with constant SF rates
over a Hubble time and a range of stellar metallicities, and the
resulting colours and M/L ratios were compared with those observed for
galaxies in the present sample.

Photometry in the $R$ band was taken from the \Ha GS database, and $J$ and
$K$ total magnitudes were derived from the 2-Micron All-Sky Survey
\citep{jarr03} as explained in section \ref{sec:data}.  The mean
$(R-K)$ and $(J-K)$ colours for the 21 Sm and Im galaxies from the
present sample with available 2MASS data are listed in the final
column of Table \ref{tbl:colours}.  The BaSTI model colours for 
stellar metallicities of [Fe/H] $= -$0.96 and $-$0.25 are listed 
in the second and third columns respectively. 

The observed mean colours appear to agree best with 
those for a model [Fe/H] slightly higher than
$-$0.96, i.e. similar to that generally found for the
Small Magellanic Cloud ([Fe/H]$\simeq$--0.7).  The M/L ratio of this same
population is almost identical to that inferred earlier using the
model of \citet{bell01}, hence there is no inconsistency in using the
mass from the latter model in calculating the SF timescale in section
\ref{sec:sft}.

\begin{table*}
\begin{center} 
\begin{tabular}{c|cc|c}
\hline
\hline
 Colour      & [Fe/H]$=$--0.96 & [Fe/H]$=$--0.25 &  Observed    \cr
\hline
 $R-K$  & 1.84 & 2.09 & 1.865$\pm$0.103 \cr
 $J-K$  & 0.76 & 0.86 & 0.777$\pm$0.018 \cr
\hline
\end{tabular}
\caption[]{Modelled and observed stellar population colours.}
\label{tbl:colours}
\end{center}
\end{table*}

\begin{figure}
\centering
\rotatebox{0}{
\includegraphics[height=7.0cm]{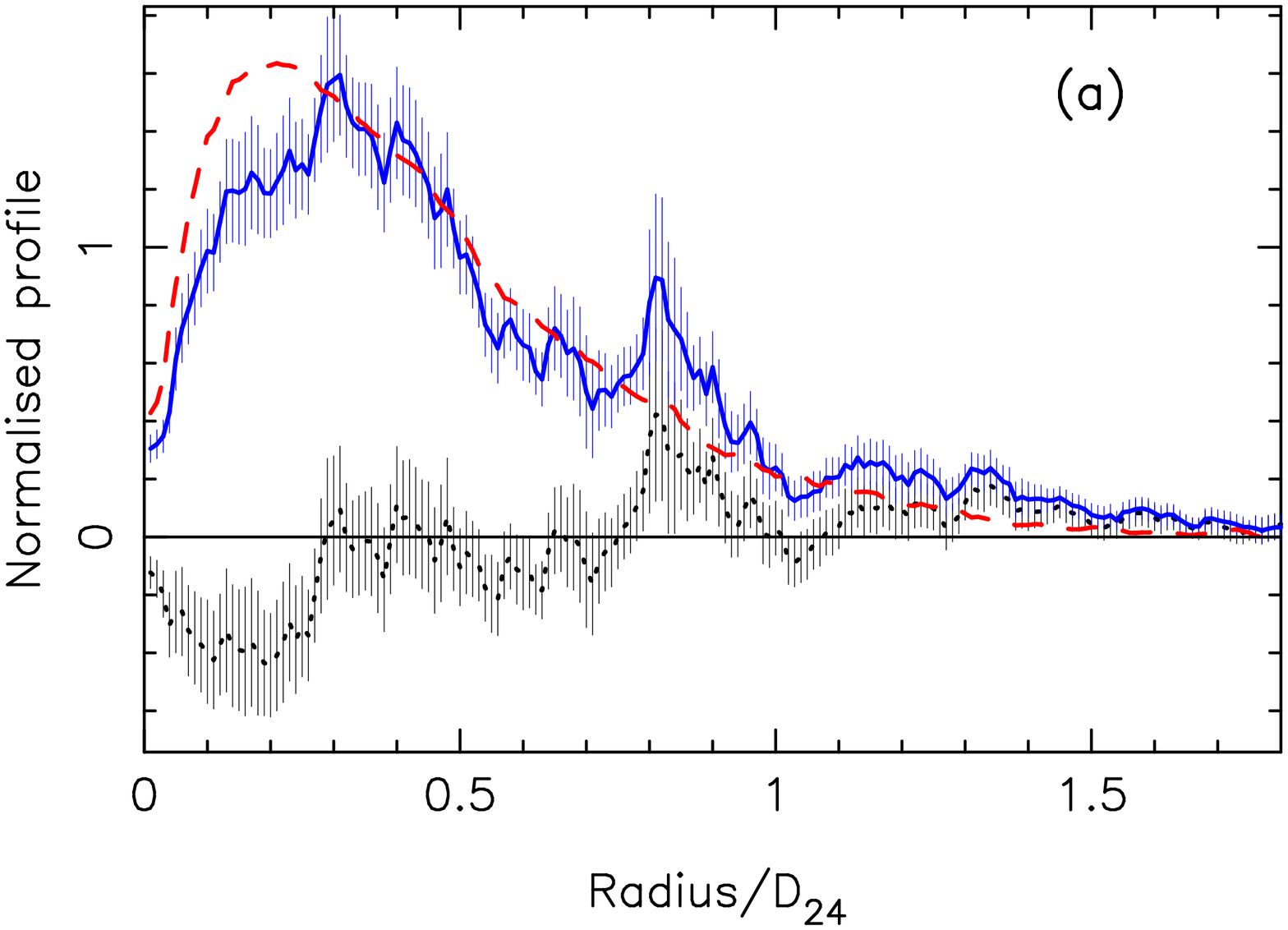}
}
\rotatebox{0}{
\includegraphics[height=7.0cm]{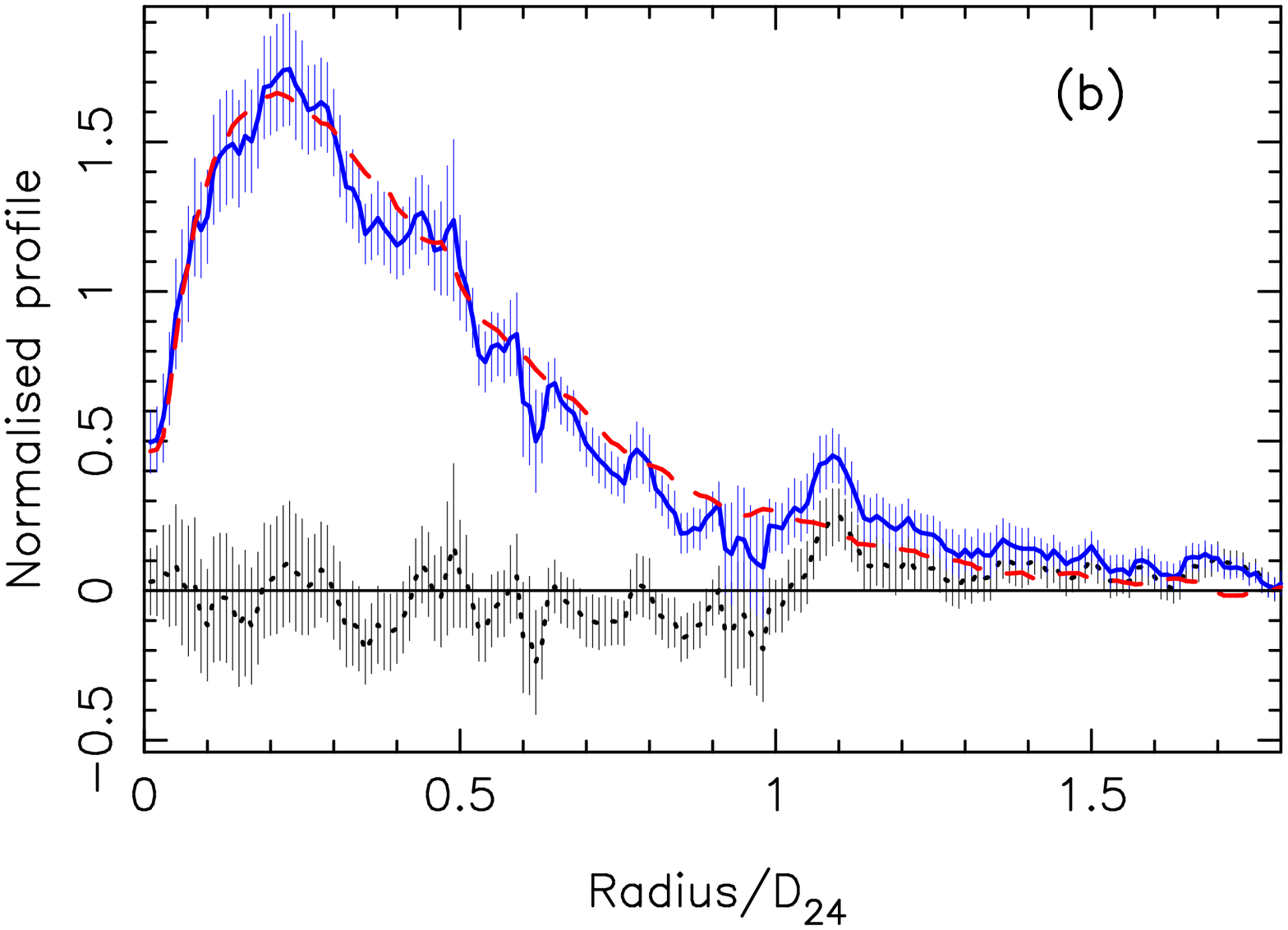}
}
\caption{(a) Mean, normalized radial distributions of \Ha\ (solid line)
and $R$-band (dashed line) light for 42 Sm galaxies; the dotted
line shows the residuals between these two profiles.
(b) The same mean profiles for the 75 Im galaxies}
\label{fig:smimprof}
\end{figure}

We can also test whether the spatial distribution of newly-formed and
old stellar populations is consistent in these galaxies.
Figure~\ref{fig:smimprof} shows the normalised distributions of the
mean $R$-band (dashed) and \Ha\ (solid line) fluxes of all Sm (a) and
Im (b) galaxies in the current sample.  The vertical scale is flux
within circular annuli ({\em not} surface brightness), normalised such
that the total area under the curve, which represents the total flux,
is unity for each galaxy.  The horizontal scale is annular radius in
units of the diameter of the galaxy at an $R$-band surface brightness
of 24~mag/sq. arcsec.  Thus this is a dimensionless shape function
describing the averaged light distributions of all the Im galaxies,
with all galaxies, large and small, receiving equal weight.  Given the
small number of \HII\ regions in many of these galaxies, the
individual galaxy \Ha\ profiles are very spiky.  In the mean, however,
they look similar to the mean $R$-band light distribution, as is
reflected by the small size of the residuals between the 2 profiles,
in the sense (${\rm H}\alpha - R$), shown by the dotted line in
Fig.~\ref{fig:smimprof}.  If we take the $R$-band light to be a fair
tracer of the distribution of the old stellar population, this is
consistent with the overall stellar population having been produced by
SF activity distributed like that observed at present.

\section{Summary}
\label{sec:sum}

We have studied the SF histories of 117 late-type
Sm and Im field galaxies from the \Ha GS sample, using the variations
in current properties to constrain the likely SF histories of 
galaxies of these types.  The main conclusions are as follows:

\begin{itemize}

\item When averaged across the whole sample, the total stellar mass 
in these galaxies could have been formed by the extrapolation of the
current SF rate over a Hubble time.  This conclusion applies to Sm and Im 
galaxies across the full mass range studied, 10$^7$ - 10$^{9.5}$~M$_{\odot}$.
This confirms the conclusions of e.g. \citet{hunt85} and \citet{vanz01},
but for a larger and more nearly complete sample.

\item Disk galaxies of types S0a - Sb, and Sbc - Sdm, have current SF
rates too low to accumulate their total stellar masses over a Hubble
time, with the deficiency being most marked for the earliest types and
largest masses.  There is substantial overlap in the stellar mass
ranges of these earlier types with the Sm/Im galaxies, so it appears
to be the presence or absence of a bulge component that determines
whether the mean SF timescale is greater than or equal to a Hubble
time.

\item Gas depletion timescales for the late-type galaxies are long
compared with a Hubble time, and longer than those of earlier type
galaxies.

\item No completely quiescent Sm or Im galaxies are found in our sample;
such galaxies must be very rare in the field population generally.

\item An analysis of the correlation in the scatter of SF and gas depletion 
times implies a typical variation in the SF rate of
a factor of 1.8 about the mean value.

\item Population synthesis modelling assuming a constant SF rate
accurately reproduces the mean colours and stellar M/L
ratios found for the Sm and Im galaxies in the present sample.

\item  The mean spatial distribution of SF in the Sm and Im galaxies
is consistent with the distribution of the old stellar population 
as seen in the $R$-band light.

\end{itemize}

\begin{acknowledgements}
MP acknowledges STFC for a postgraduate studentship.
The Jacobus Kapteyn Telescope was operated on the island of La Palma
by the Isaac Newton Group in the Spanish Observatorio del Roque de los
Muchachos of the Instituto de Astrof\'isica de Canarias. This research
has made use of the NASA/IPAC Extragalactic Database (NED) which is
operated by the Jet Propulsion Laboratory, California Institute of
Technology, under contract with the National Aeronautics and Space
Administration.  The referee is thanked for many helpful suggestions,
which improved the content, clarity and presentation of the paper. 
\end{acknowledgements}

\bibliographystyle{bibtex/aa}
\bibliography{refs}
      

\end{document}